\documentstyle[preprint,aps]{revtex}

\title{\vspace*{78pt} \Large \bf
       A new approach to the integration of rotational
       motion in systems with interacting rigid bodies \\ [24pt]}

\author{\large Igor P. Omelyan \\ [36pt]}

\address{\normalsize \em
         Institute for Condensed Matter Physics,
         National Ukrainian Academy of Sciences, \\
         1 Svientsitsky st., UA-290011 Lviv, Ukraine.
         E-mail: nep@icmp.lviv.ua \\ [64pt]}

\date{\today}

\begin{document}

\maketitle

\vspace{6.5cm}



\newpage

\vspace*{48pt}

\begin{center}

{\em A running head:}
\large

Numerical integration of rigid-body motion

\vspace{80pt}

\begin{tabular}{l}
Dr. I. P. Omelyan \\
Institute for Condensed Matter Physics \\
National Ukrainian Academy of Sciences \\
1 Svientsitsky st., UA-290011 Lviv \\
UKRAINE \\ [24pt]
Tel/Fax: +380--322--761978 \\
E-mail: nep@icmp.lviv.ua

\end{tabular}
\end{center}

\newpage

\vspace*{60pt}

\begin{abstract}

A new approach is developed to integrate numerically the equations
of motion for systems of interacting rigid polyatomic molecules. With
the aid of a leapfrog framework, we directly involve principal angular
velocities into the integration, whereas orientational positions are
expressed in terms of either principal axes or quaternions. As a result,
the rigidness of molecules appears to be an integral of motion, despite
the atom trajectories are evaluated approximately. The algorithm derived
is free of any iterative procedures and it allows to perform both energy-
and temperature-conserving simulations. The corresponding integrators are
time reversible but the symplectic behavior is only achieved in mean.
Symplectic versions are also described. They provide the conservation
of volume in phase space precisely at each time step and, moreover, lead
to exact solutions for angular velocities in the inertial-motion regime.
It is shown that the algorithm exhibits excellent stability properties
and conserves the energy even somewhat better than the atomic-constraint
technique.

\end{abstract}

\newpage

\section{Introduction}

A lot of theories in physical chemistry treats the liquid as a system of
classical rigid bodies. The method of molecular dynamics (MD) remains the
main tool for investigation of such a model. An important problem in MD
simulations is the development of stable and efficient algorithms for
integrating the equations of motion with orientational degrees of freedom.
The straightforward parameterization of these degrees, Euler angles, is
very inefficient for numerical calculations because of singularities
inherent in the description [1--3]. Within singularity free approaches,
the orientations of molecules typically are expressed in terms of
quaternions [4--6]. High-order Gear methods were utilized to integrate
the rigid-body equations of motion in early investigations [7--10].
These schemes are not reversible or symplectic, and it is not clear
that the extra order obtained is relevant, since they quickly become
exhibit poor long-term stability of energy with increasing the step
size [5, 11].

Various alternatives to the Gear approach have been proposed and implemented
over the years. These include Verlet [12], velocity Verlet [13], leapfrog
[14], and Beeman [15] integrators, which are based on the St\"ormer central
difference approximation [7, 16] of accelerations. Such integrators proved
to be the most efficient, because high accuracy can be reached with minimal
cost measured in terms of force evaluations. The main problem with these
methods is that they were initially constructed, in fact, to integrate
translational motion assuming the velocity-independence of the
accelerations, and, therefore, additional modifications are
necessary to use them for rotational dynamics.

In the atomic approach [17--21], the parameterization of orientational
degrees of freedom is circumvented by involving individual Cartesian
coordinates of atomic sites. As a result, the Verlet algorithm and its
velocity versions appear to be directly applicable. The dynamics is
determined by integrating the equations of motion for these sites,
subject to constraints that the intramolecular bond distances are fixed.
Until now, the great majority of MD simulations on polyatomics are
performed using the atomic-constraint technique due to its exceptional
long-time stability. Although this approach can be applied, in principle,
to arbitrary systems including the case of flexible molecules, it has
some disadvantages. For example, to exactly reproduce the rigid molecular
structure, complicated systems of nonlinear equations are needed to be
solved by iterations at each time step of the integration process. In
general, the convergence is not guaranteed [22] and looping becomes
possible already at relatively small step sizes, especially for
macromolecules with bond angle reactions. An additional complexity
arises for point molecules with embedded multipoles, since then the
forces are not easily decomposable into direct site-site contributions.

In order to obviate these difficulties, Ahlrichs and Brode have devised a
hybrid method [23] in which the principal axes of molecules are considered
as pseudo particles and constraint forces are introduced to maintain their
orthonormality. The principal axes were evaluated within the Verlet
framework via a recursive procedure which does not solve exactly the
constraint equations to convergence, but instead writes the rotational
matrix as an exponential for sum of some anti-symmetric matrices,
restricting by a finite number of terms. It was established, however,
that the exponential propagation leads to much worse results on energy
conservation than those obtained in the atomic-constraint technique.
Recently, acting in the spirit of the pseudo-particle formalism, a
leapfrog-like algorithm has also been proposed [24]. In this algorithm
the entire rotation matrix and the corresponding conjugate momentum are
treated as dynamical variables, and the matrix of constrain forces is
evaluated exactly. In such a way, the lost precision was reproduced, but
this required again to find iterative solutions to highly nonlinear
equations. Moreover, since velocities do not appear explicitly, it is
hard to extend the pseudo-particle approach to thermostatted versions.

The first efforts on adopting the St\"ormer group of integrators to
rotational motion in its pure form were done by Fincham [25--26]. As a
result, explicit and implicit angular-momentum leapfrog algorithms have
been introduced. In the case of a more accurate implicit version, the
system of four nonlinear equations per molecule is solved iteratively
for the same number of quaternion components [26]. As was soon realized
[27], the Fincham's algorithms are not very efficient in energy-conserving
simulations, given that the artificial rescaling procedure is used to
maintain the unit quaternion norm, and additional transformations with
approximately computed rotational matrices and angular momenta are
necessary to evaluate the principal components of angular velocities.
The question of how to replace the crude renormalization by a more natural
procedure has been considered too [28]. As a consequence, the quaternion
dynamics with constraints was formulated within the velocity Verlet
framework. Similar ideas were used to adopt the pseudo-particle approach
to velocity-Verlet and Beeman integrators [27]. It has been shown that
such algorithms conserve the total energy better with respect to the
implicit leapfrog integrator [26] but worse than the atomic-constraint
method, especially in the case of long-duration simulations with large
step sizes.

Among other techniques developed in recent years it is necessary to mention
the multiple time scale (MTS) integration [29-34]. The reversible reference
system propagator (RESP) algorithm by Berne {\em at al.} [32] is a general
approach which yields a whole family of MTS integrators. The basic idea of
this approach lies in using a short time step for the rapidly varying fast
forces, and a large time step for the nonbonding slow forces arising
from the interactions at long interparticle distances. Then the most
time-consuming slow forces will be recalculated less frequently than
in usual methods, saving significantly the computer time. It is worth
remarking that the RESP approach, being originally constructed on
Cartesian coordinates, can be adapted [35] to orientational variables.
There are some arguments to stay that a reformulation of the RESP approach
within these variables can be more convenient for applications (see Sec.
IV). From this point of view, and taking into account that the fastest
motions are handled (inside the innermost cycle with the least step size)
using standard algorithms, the developing of methods for direct integration
of orientational variables presents an interest in the context of the MTS
approach as well.

Quite recently, to improve the efficiency of integration in orientational
space, a new angular-velocity leapfrog algorithm has been proposed [36].
Its main advantages were the intrinsic conservation of rigid structures
and high stability properties. However, a common drawback, inherent in
all precise long-term integrators on rigid polyatomics, still remained,
namely, the necessity to solve by iteration nonlinear equations.

In the present paper we develop the angular-velocity algorithm by
avoiding any iterative procedures. A thermostatted version is also
introduced. We demonstrate that the integrators derived are time
reversible, symplectic and the total energy is conserved even better
than within the atomic-constraint method. It is discussed that our
approach can be especially useful for simulations of systems at high
temperatures, where the inertial-motion regime plays an important role
in rotational dynamics.

\section{Advanced leapfrog algorithm}

We shall deal with a classical collection of $N$ rigid molecules composed
of $M$ point interaction sites. According to the molecular approach, the
dynamics for such a system can be considered in view of translational and
rotational motions. The translational motion is expressed in terms of the
center of mass ${\bf r}_i=\sum_{a=1}^M m_a {\bf r}_i^a/m$ of each molecule
$(i=1,\ldots,N)$, where ${\bf r}_i^a$ denotes the positions of atomic site
$a$ within molecule $i$ and $m=\sum_{a=1}^M m_a$ and $m_a$ are the masses of
a separate molecule and partial atoms, respectively. Then the translational
part of time evolution can be determined by the first-order differential
equations: ${\rm d} {\bf r}_i/{\rm d} t = {\bf v}_i$ and $m {\rm d} {\bf
v}_i/{\rm d} t = {\bf f}_i$, where ${\bf f}_i=\sum_{j;a,b}^{N;M} {\bf
f}_{ij}^{ab} (|{\bf r}_i^a-{\bf r}_j^b|)$ is the force acting on molecule
$i$ due to the site-site interactions ${\bf f}_{ij}^{ab}$ with the other
($j \ne i$) molecules and ${\bf v}_i$ designates the center-of-mass
velocity.

\subsection{Rotational motion in body-vector and quaternion representations}

In the body-vector representation [4, 23, 27], the Cartesian coordinates of
three principal axes $XYZ$ of the molecule are assumed to be orientational
variables. These variables can be collected into the $3 \times 3$ orthonormal
matrices ${\bf A}_i$, so that the site positions in the laboratory frame are:
${\bf r}_i^a(t)={\bf r}_i(t)+{\bf A}_i^+(t) {\bf \Delta}_a$, where ${\bf
\Delta}_a=(\Delta^a_X, \Delta^a_Y, \Delta^a_Z)^+$ is a vector-column of
these positions in the body frame attached to the molecule and ${\bf A}^+$
denotes the matrix transposed to ${\bf A}$. The time-independent quantities
${\bf \Delta}_a$ ($a=1,\ldots,M$) are completely defined by the rigid
molecular structure. The rate of change in time of orientational matrices
can be given as
\begin{equation}
\frac{{\rm d} {\bf A}_i}{{\rm d} t}=\left(
\begin{array}{ccc}
0 & \Omega_Z^i & -\Omega_Y^i \\
-\Omega_Z^i & 0 & \Omega_X^i \\
\Omega_Y^i & -\Omega_X^i & 0
\end{array}
\right)
{\bf A}_i \equiv {\bf W}({\bf \Omega}_i) {\bf A}_i \, ,
\end{equation}
where $\Omega_X^i$, $\Omega_Y^i$ and $\Omega_Z^i$ are principal components
of the angular velocity ${\bf \Omega}_i$, and ${\bf W}$ is a skewsymmetric
matrix associated with ${\bf \Omega}_i$, i.e., ${\bf W}^+({\bf \Omega}_i)=
-{\bf W}({\bf \Omega}_i)$.

In an alternative approach [4, 9], the matrix ${\bf A}_i$ is a function,
\begin{equation}
{\bf A} ({\bf q}_i) =
\left( \begin{array}{ccc}
-\xi_i^2+\eta_i^2-\zeta_i^2+\chi_i^2 & 2(\zeta_i \chi_i - \xi_i \eta_i) &
2 (\eta_i \zeta_i + \xi_i \chi_i) \\
-2(\xi_i \eta_i + \zeta_i \chi_i) & \xi_i^2-\eta_i^2-\zeta_i^2+\chi_i^2 &
2(\eta_i \chi_i - \xi_i \zeta_i) \\
2(\eta_i \zeta_i - \xi_i \chi_i) & -2(\xi_i \zeta_i + \eta_i \chi_i) &
-\xi_i^2-\eta_i^2+\zeta_i^2+\chi_i^2
\end{array}
\right) \, ,
\end{equation}
of the four-component quaternion ${\bf q}_i \equiv (\xi_i,\eta_i,\zeta_i,
\chi_i)^+$. The time derivatives of ${\bf q}_i$ can be presented [36] in
the form
\begin{equation}
\frac{{\rm d} {\bf q}_i}{{\rm d}t}
= \displaystyle \frac12 \left(
\begin{array}{cccc}
0&\Omega_Z^i&-\Omega_X^i&-\Omega_Y^i \\
-\Omega_Z^i&0&-\Omega_Y^i&\Omega_X^i\\
\Omega_X^i&\Omega_Y^i&0&\Omega_Z^i  \\
\Omega_Y^i&-\Omega_X^i&-\Omega_Z^i&0
\end{array}
\right)
\left( \begin{array}{c}
\xi_i \\ \eta_i \\ \zeta_i \\ \chi_i
\end{array} \right)
\equiv \displaystyle \frac12 {\bf Q}({\bf \Omega}_i) {\bf q}_i \, ,
\end{equation}
where ${\bf Q}({\bf \Omega}_i)$ is a skewsymmetric matrix again. The
normalization condition ${{\bf q}_i}^2=\xi_i^2+\eta_i^2+\zeta_i^2+\chi_i^2=
1$, which ensures the orthonormality ${\bf A}_i {\bf A}_i^+={\bf I}$ of
${\bf A}_i \equiv {\bf A} ({\bf q}_i)$, where ${\bf I}$ designates the
unit matrix, has been used to obtain Eq.~(3).

The relations (1) and (3) for orientational coordinates need to be
supplemented by the Euler's equations for angular velocities
\vspace{2pt}
\begin{eqnarray}
J_X \frac{{\rm d} {\Omega}_X^i}{{\rm d} t}&=&K_X^i
+ \left( J_Y-J_Z \right) {\Omega}_Y^i {\Omega}_Z^i \, , \nonumber \\
J_Y \frac{{\rm d} {\Omega}_Y^i}{{\rm d} t}&=&K_Y^i
+ \left( J_Z-J_X \right) {\Omega}_Z^i {\Omega}_X^i \, , \\
J_Z \frac{{\rm d} {\Omega}_Z^i}{{\rm d} t}&=&K_Z^i
+ \left( J_X-J_Y \right) {\Omega}_X^i {\Omega}_Y^i \, , \nonumber
\end{eqnarray}
where $J_\alpha$ denote the time-independent moments inertia of the molecule
along its principal axes ($\alpha=X,Y,Z$) and $K_\alpha^i$ are body-frame
components, ${\bf K}_i={\bf A}_i {\bf k}_i$, of the torque ${\bf k}_i=
\sum_{j;a,b}^{N;M} ({\bf r}_i^a-{\bf r}_i) {\mbox{\boldmath $\times$}}
{\bf f}_{ij}^{ab}$ exerted on molecule $i$ with respect to its center
of mass.

\subsection{Evaluation of angular velocities and coordinates}

Let $\{{\bf r}_i(t), {\bf v}_i(t-{\textstyle \frac{h}{2}}), {\bf
S}_i(t), {\bf \Omega}_i(t-{\textstyle \frac{h}{2}})\}$ be an initial
spatially-velocity configuration of the system, where the velocities
and positions are defined on alternate half-time steps with $h$ being
the fixed step size and ${\bf S}_i(t) \equiv {\bf A}_i(t)$ or ${\bf
q}_i(t)$ are the orientational coordinates for the principal-axis or
quaternion representations, respectively. The translational variables
can be integrated applying the usual [14] leapfrog algorithm
\begin{eqnarray}
{\bf v}_i(t+{\textstyle \frac{h}{2}})&=&
{\bf v}_i(t-{\textstyle \frac{h}{2}})+
h {\bf f}_i(t)/m + {\cal O}(h^3) \, , \nonumber \\ [-12pt] \\ [-12pt]
{\bf r}_i(t+h)&=&{\bf r}_i(t)+h {\bf v}_i(t+{\textstyle \frac{h}{2}})
+ {\cal O}(h^3) \nonumber
\end{eqnarray}
in which forces ${\bf f}_i(t)$ are explicitly evaluated in terms of known
spatial coordinates ${\bf r}_i(t)$ and ${\bf S}_i(t)$. As can be verified
easily, expanding the left- and right-hand sides of both the lines of
Eq.~(5) into Taylor series over $h$, the algorithm produces truncation
single-step errors of order $h^3$ in coordinates and velocities.

In the case of rotational motion it is not a simple matter to adopt the
standard leapfrog scheme, since the principal angular accelerations are
velocity-dependent and the time derivatives of orientational coordinates
depend not only on the angular velocity but also on the coordinates
themselves. These problems were handled previously by Fincham [26] in his
angular-momentum versions of the leapfrog algorithm. Recently [27, 36], it
was shown that more superior techniques follow when principal angular
velocities are involved directly into the integration. Within the leapfrog
framework, mid-step values for the angular velocities can be evaluated by
writing
\begin{equation}
{{\Omega}_\alpha^i}(t+{\textstyle \frac{h}{2}})=
{\Omega}_\alpha^i(t-{\textstyle \frac{h}{2}})+
h \Big[ K_\alpha^i(t) + \left( J_\beta-J_\gamma \right)
{{\Omega}_\beta^i}(t) {{\Omega}_\gamma^i}(t) \Big]
\Big/ {J_\alpha} + {\cal O}(h^3) \, ,
\end{equation}
where Euler equations (4) have been taken into account, $(\alpha,\beta,
\gamma)$ denote an array of three cyclic permutations for $(X,Y,Z)$, and
torques $K_\alpha^i(t)$ are computed via coordinates ${\bf r}_i(t)$ and
${\bf S}_i(t)$. Equation (6) presents a rotational-motion analog for the
first line of (5) but it must be complemented by an interpolation of the
products of angular velocities to on-step levels of time. It is quite
naturally to perform such an interpolation in the form
\begin{equation}
{{\Omega}_\beta^i}(t) {{\Omega}_\gamma^i}(t) =
\frac12 \left[ {\Omega}_\beta^i(t-{\textstyle \frac{h}{2}})
{\Omega}_\gamma^i(t-{\textstyle \frac{h}{2}})+
{{\Omega}_\beta^i}(t+{\textstyle \frac{h}{2}})
{{\Omega}_\gamma^i}(t+{\textstyle \frac{h}{2}}) \right] + {\cal O}(h^2) \, ,
\end{equation}
where ${\cal O}(h^2)$ uncertainties are in the self-consistency with the
second-order accuracy of angular-velocity propagation (6). In view of (7),
vector expression (6) is an implicit equation with respect to ${\bf
\Omega}_i(t+{\textstyle \frac{h}{2}})$, which allows to be solved by
iteration [36].

Similarly to the translational-position evaluations (second line of
Eq.~(5)), we integrate orientational coordinates,
\begin{equation}
{\bf S}_i(t+h)={\bf S}_i(t) + h {\bf H}_i(t+{\textstyle \frac{h}{2}})
{\bf S}_i(t+{\textstyle \frac{h}{2}}) + {\cal O}(h^3) \, ,
\end{equation}
for the cases of principal-axis vectors $({\bf S}_i \equiv {\bf A}_i,
{\bf H}_i \equiv {\bf W}_i)$ and quaternion $({\bf S}_i \equiv {\bf q}_i,
{\bf H}_i \equiv \frac12 {\bf Q}_i)$ representations, where Eqs. (1) and
(3) have been used. The matrices ${\bf W}_i \equiv {\bf W} ({\bf \Omega}_i)$
and ${\bf Q}_i \equiv {\bf Q}({\bf \Omega}_i)$ are calculated in (8) using
the already defined values of ${\bf \Omega}_i(t+{\textstyle \frac{h}{2}})$,
whereas the obvious choice for mid-step values of orientational variables is
\begin{equation}
{\bf S}_i(t+{\textstyle \frac{h}{2}})={\textstyle \frac12}
\left[{\bf S}_i(t)+{\bf S}_i(t+h)\right] + {\cal O}(h^2) \, .
\end{equation}
Equations (8) and (9) constitute, in fact, a system of linear equations
with respect to elements of ${\bf A}_i(t+h)$ or ${\bf q}_i(t+h)$, which,
is solved analytically,
\begin{equation}
{\bf S}_i(t+h)=
[{\bf I}-{\textstyle \frac{h}{2}} {\bf H}_i(t+{\textstyle \frac{h}{2}})]^{-1}
[{\bf I}+{\textstyle \frac{h}{2}} {\bf H}_i(t+{\textstyle \frac{h}{2}})]
{\bf S}_i(t) + {\cal O}(h^3) \, ,
\end{equation}
where it is understood that ${\bf I}$ designates either three- or
four-dimensional unit matrix in the principal-axis or quaternion
domains, respectively.

Taking into account expressions (1) and (3) for matrix ${\bf H}_i$, the
result (10) can be written more explicitly,
\vspace{3pt}
\begin{eqnarray}
\textstyle
{\bf A}_i(t+h)&=&\frac{{\bf I}\,[1-{\textstyle \frac{h^2}{4}}
\Omega_i^2(t+{\textstyle \frac{h}{2}})]+h{\bf W}_i+
{\textstyle \frac{h^2}{2}}{\bf P}_i}{1+
{\textstyle \frac{h^2}{4}} \Omega_i^2(t+{\textstyle \frac{h}{2}})} \,
{\bf A}_i(t) \equiv {\bf D}_i(t,h) \, {\bf A}_i(t) \, ,
\nonumber \\ [-5pt] \\ [-5pt]
{\bf q}_i(t+h)&=&\frac{{\bf I}\,[1-{\textstyle \frac{h^2}{16}}
\Omega_i^2(t+{\textstyle \frac{h}{2}})]+\frac{h}{2}
{\bf Q}_i}{1+{\textstyle \frac{h^2}{16}}
\Omega_i^2(t+{\textstyle \frac{h}{2}})}
\, {\bf q}_i(t) \equiv {\bf G}_i(t,h) \, {\bf q}_i(t) \, ,
\nonumber
\end{eqnarray}

\vspace{3pt}

\noindent
where $[{\bf P}_i]_{\alpha \beta}={\Omega}_\alpha^i {\Omega}_\beta^i$
denotes a symmetric matrix which, as the matrices ${\bf W}_i$ and
${\bf Q}_i$, is computed in (11) using the angular velocities at
middle-step time $t+{\textstyle \frac{h}{2}}$. It can be checked that
the matrix $({\bf I}-\epsilon {\bf H})^{-1} ({\bf I}+\epsilon {\bf H})$
is orthonormal at arbitrary values of $\epsilon$, provided ${\bf H}^+=
-{\bf H}$. As a result, the $3 \times 3$ and $4 \times 4$ evolution
matrices ${\bf D}_i$ and ${\bf G}_i$ are orthonormal by construction
as well. Using the equalities ${\bf W}_i^2={\bf P}_i-\Omega_i^2 {\bf I}$
and ${\bf Q}_i^2=-\Omega_i^2 {\bf I}$, these matrices can be presented
in the exclusive compact form
\begin{equation}
{\bf D}_i(t,h) = {\rm \bf exp}[\varphi_i {\bf W}_i/
\Omega_i]_{t+\frac{h}{2}} \, , \ \ \ \ \
{\bf G}_i(t,h) = {\rm \bf exp}[\phi_i {\bf Q}_i/
\Omega_i]_{t+\frac{h}{2}} \, ,
\end{equation}
where $\varphi_i=\arcsin [h \Omega_i(t+{\textstyle \frac{h}{2}})/(1+
{\textstyle \frac{h^2}{4}} \Omega_i^2(t+{\textstyle \frac{h}{2}}))]$,
$\phi_i=\arcsin [h \Omega_i(t+{\textstyle \frac{h}{2}})/(2+{\textstyle
\frac{h^2}{8}} \Omega_i^2(t+{\textstyle \frac{h}{2}}))]$ and ${\rm \bf
exp}$ designates the matrix exponential. Then it becomes clear that
matrices ${\bf D}_i$ and ${\bf G}_i$ define three- and four-dimensional
rotations on angles $\varphi_i$ and $\phi_i$ in the laboratory frame and
quaternion space, respectively. In the first case the rotation is performed
around the unit vector ${\bf \Omega}_i/\Omega_i|_{t+\frac{h}{2}}$, whereas
in the second one it is carried out around a virtual orth which is
perpendicular to all four orths of the quaternion space.

From the above, the following important statement emerges immediately.
If initially the orthonormality of ${\bf A}_i$ and the unit norm of
${\bf q}_i$ are fulfilled, they will be satisfied perfectly in future,
despite an approximate character of the evaluated trajectories. This
excellent property distinguishes the algorithm introduced from all other
singularity-free algorithms known, since no artificial or constraint
normalizations as well as no recursive or iterative procedures are
necessary to conserve the rigidness of molecules.

\subsection{Thermostatted dynamics}

Since the velocities appear explicitly in our approach, it is possible to
introduce various thermostats [37-43] to simulate the canonical ensemble.
Usually [26, 44], thermostatted versions allow to perform simulations with
greater step sizes than those used within the energy-conserving dynamics.
In canonical MD simulations the time evolution should be determined in such
a way to keep a fixed temperature $T = \langle \cal T \rangle$ of the system,
where ${\cal T} = 2 {\mit \Gamma}/(l N k_{\rm B})$ and ${\mit \Gamma}$ are
the instantaneous temperature and kinetic energy, respectively, $k_{\rm B}$
is the Boltzmann's constant, and $\langle \ \rangle$ denotes the statistical
averaging. The number $l$ of degrees of freedom per particle is equal to six
for arbitrary rigid polyatomics, except linear molecules when $l=5$.

We shall consider here a N\'ose-Hoover thermostat [39--41], although
the extension to others thermostats [37, 38, 43] can also be realized.
According to the N\'ose-Hoover technique, the thermostatted dynamics is
obtained introducing the generalized friction forces $-\lambda m {\bf v}_i
\equiv -\lambda \partial {\mit \Gamma}/\partial {\bf v}_i$. These virtual
forces are added to the real ones and, as a result, the equations of motion
for translational velocities transform into $m {\rm d} {\bf v}_i/{\rm d} t
= {\bf f}_i - \lambda m {\bf v}_i$. In our case, when orientational degrees
of freedom are present additionally, the kinetic energy consists of both
translational and rotational parts, ${\mit \Gamma}=\frac12 \sum_{i=1}^N [m
{{\bf v}_i}^2 + \sum_{\alpha}^{X,Y,Z} J_\alpha {\Omega_\alpha^i}^2]$. This
requires the introduction of friction torques, $-\lambda \partial {\mit
\Gamma}/\partial \Omega_\alpha^i=-\lambda J_\alpha \Omega_\alpha^i$, which
should be taken into account in Euler equations (4). The time-dependent
friction coefficient varies in time with the total excess of kinetic energy
to its canonical mean value,
\vspace{-6pt}
\begin{equation}
{\rm d} \lambda / {\rm d} t = ({\cal T} - T)/(T \tau^2) \, ,
\end{equation}

\vspace{-6pt}

\noindent
and governs by a characteristic thermostat relaxation time $\tau$.

In the presence of friction forces, the translational velocities can be
integrated applying a Toxvaerd leapfrog algorithm [45]. It is based on
the estimation
\begin{equation}
{\bf v}_i(t)={\textstyle \frac12}
[{\bf v}_i(t-{\textstyle \frac{h}{2}})+
{\bf v}_i(t+{\textstyle \frac{h}{2}})] + {\cal O}(h^2)
\end{equation}
of on-step velocities, commonly used to calculate a kinetic part of the
total energy at time $t$ in microcanonical simulations and to verify
the energy conservation. Then adding the corresponding friction term
$\lambda(t) m {\bf v}_i(t)$ to the right-hand side of the first line
of Eq.~(5) and solving the obtained equation with respect to new
mid-step translational velocities, one finds
\begin{equation}
{\bf v}_i(t+{\textstyle \frac{h}{2}})=
\{ [1-{\textstyle \frac{h}{2}} \lambda(t)]
{\bf v}_i(t-{\textstyle \frac{h}{2}})+
h {\bf f}_i(t)/m \}/ [1+{\textstyle \frac{h}{2}} \lambda(t)] \, .
\end{equation}

\vspace{-4pt}

\noindent
Quite analogously, in the case of rotational motion we use the estimation
\vspace{-4pt}
\begin{equation}
{\bf \Omega}_i(t)={\textstyle \frac12}
[{\bf \Omega}_i(t-{\textstyle \frac{h}{2}})+
{\bf \Omega}_i(t+{\textstyle \frac{h}{2}})] + {\cal O}(h^2) \, ,
\end{equation}

\vspace{-4pt}

\noindent
for on-step angular velocities, needed to computer the friction torques
$-\lambda(t) J_\alpha \Omega_\alpha^i(t)$. These torques we add to the
right-hand side of Eq.~(6) which now can be presented in the following
form
\vspace{-4pt}
\begin{equation}
\Omega_\alpha^i(t+{\textstyle \frac{h}{2}})=
\{ [1-{\textstyle \frac{h}{2}} \lambda(t)]
\Omega_\alpha^i(t-{\textstyle \frac{h}{2}})+
h {\cal K}_\alpha^i(t)/J_\alpha \}/
[1+{\textstyle \frac{h}{2}} \lambda(t)] \, ,
\end{equation}

\vspace{-12pt}

\noindent
where
\vspace{-6pt}
\begin{equation}
{\cal K}_\alpha^i(t)=K_\alpha^i(t) + {\textstyle \frac12}
\left( J_\beta-J_\gamma \right)
\Big[ {\Omega}_\beta^i(t-{\textstyle \frac{h}{2}})
{\Omega}_\gamma^i(t-{\textstyle \frac{h}{2}})+
{{\Omega}_\beta^i}(t+{\textstyle \frac{h}{2}})
{{\Omega}_\gamma^i}(t+{\textstyle \frac{h}{2}}) \Big] \, .
\end{equation}
The vector equation (17) constitutes a system of three equations per
molecule for the same number of unknowns $\Omega_\alpha^i(t+{\textstyle
\frac{h}{2}})$ (note that $\lambda(t)$ is known from the previous time
step). As in the case of microcanonical evaluation ($\lambda(t) \equiv
0$), the system can be solved iteratively, replacing initially ${\bf
\Omega}_i(t+{\textstyle \frac{h}{2}})$ by ${\bf \Omega}_i(t-{\textstyle
\frac{h}{2}})$ in all nonlinear terms which are collected in the right-hand
side of (17). The calculated values of ${\bf \Omega}_i(t+{\textstyle \frac
{h}{2}})$ are then considered as initial guess for the next iteration. The
convergence of iterations is justified by the smallness of $h$ which always
is meet in actual MD simulations.

Using the already defined translational and angular velocities, we evaluate
the instantaneous mid-step temperature ${\cal T}(t+{\textstyle \frac{h}{2}})
\equiv {\cal T}(\{{\bf v}_i(t+{\textstyle \frac{h}{2}}), {\bf \Omega}_i(t+
{\textstyle \frac{h}{2}})\})$. Then equation (13) is integrated as follows
[45]:
\vspace{-4pt}
\begin{equation}
\lambda(t+h) = \lambda(t) + h
[{\cal T}(t+{\textstyle \frac{h}{2}}) - T]/(T \tau^2) \, .
\end{equation}

\vspace{-4pt}

\noindent
The coordinates ${\bf S}_i(t+h)$ and ${\bf r}_i(t+h)$ are updated according
to the same transformations (see Eq.~(11) and the second line of Eq.~(5))
as for energy-conserving dynamics.

\subsection{Avoidance of iterative procedures}

An essential advantage of the approach presented lies in the fact that
solutions to Eq.~(17) (or (6)) can be obtained without applying any
iterative procedures. First of all it is necessary to point out that
the equations are nonlinear only when all the principal moments of inertia
are different. Let us consider now this more difficult case (specific
examples are described in Subsect.~II~F) and assume for definiteness that
$J_X < J_Y < J_Z$. Then the first two unknowns $\Omega_X(t+{\textstyle
\frac{h}{2}})$ and $\Omega_Y(t+{\textstyle \frac{h}{2}})$ are the most
fast variables which should be excluded from the iterations to increase
the convergence.

Such an excluding indeed can be carried out solving the first two ($\alpha=
X,Y$) equations of system (17) with respect to $\Omega_X^i(t+{\textstyle
\frac{h}{2}})$ and $\Omega_Y^i(t+{\textstyle \frac{h}{2}})$. The result is
\begin{equation}
\Omega_X^i(t+{\textstyle \frac{h}{2}})=
\frac{\theta_X + h \rho_X \theta_Y \Omega_Z^i(t+{\textstyle \frac{h}{2}})}
{1+h^2 \mu^2 {\Omega_Z^i}^2(t+{\textstyle \frac{h}{2}})} \, , \ \ \ \ \
\Omega_Y^i(t+{\textstyle \frac{h}{2}})=
\frac{\theta_Y + h \rho_Y \theta_X \Omega_Z^i(t+{\textstyle \frac{h}{2}})}
{1+h^2 \mu^2 {\Omega_Z^i}^2(t+{\textstyle \frac{h}{2}})} \, ,
\end{equation}
where $\theta_\alpha= \nu_- \Omega_\alpha^i(t-{\textstyle \frac{h}{2}})
/\nu_+ + [K_\alpha^i(t)/(J_\alpha \nu_+)+\rho_\alpha \Omega_\beta^i(t-
{\textstyle \frac{h}{2}}) \Omega_\gamma^i(t-{\textstyle \frac{h}{2}})] h$,
$\nu_{\pm}=1 \pm {\textstyle \frac{h}{2}} \lambda(t)$, $\rho_\alpha =
\sigma_\alpha/\nu_+$, $\sigma_\alpha =(J_\beta-J_\gamma)/(2 J_\alpha)$
and $0 < \mu^2=- \rho_X \rho_Y \le 1/(4 \nu_+^2)$. The last inequalities
follow from the requirements $J_\alpha > 0$ and $J_\alpha \le J_\beta +
J_\gamma$ imposed on principal moments of inertia. It is worth remarking
that putting formally $\lambda(t) \equiv 0$, i.e. $\nu_{\pm} \equiv 1$,
we shall come to solutions of Eq.~(6) corresponding to the microcanonical
ensemble. In view of (20), only the third equation ($\alpha=Z$) of system
(17) really needs to be iterated with respect to one variable $\Omega_Z^i
(t+{\textstyle \frac{h}{2}})$. Since this variable is the most slow
quantity, a well convergence will be guaranteed even for not so well
normally behaved case as an almost linear body, when $J_X \ll J_Y < J_Z$.

Finally, we shall show how to obviate the iterative solutions at all.
Substituting result (20) into the third equation of system (17) and
presenting the $Z$-th component of angular velocity in the form
$\Omega_Z^i(t+{\textstyle \frac{h}{2}})=s_0+\delta$ lead to the
following algebraic equation
\begin{equation}
a_0+a_1 \delta+a_2 \delta^2+a_3 \delta^3+a_4 \delta^4+a_5 \delta^5=0
\end{equation}
with the coefficients
\begin{eqnarray}
a_0&=&(s_0 - \theta_Z) \vartheta_+^2 -
h \rho_Z [ \theta_X \theta_Y \vartheta_-
+ h (\rho_Y \theta_X^2 + \rho_X \theta_Y^2) s_0 ] \, , \nonumber \\
a_1&=&\vartheta_+ - h^2 \{ (\rho_Y \theta_X^2 + \rho_X \theta_Y^2) \rho_Z
- \mu^2 s_0 [(5 s_0 - 4 \theta_Z) \vartheta_+
+ 2 h \theta_X \theta_Y  \rho_Z ] \} \, ,\nonumber \\
a_2&=&h^2 \mu^2 [6 s_0-2 \theta_Z + h \rho_Z \theta_X \theta_Y
+ h^2 \mu^2 s_0^2 (10 s_0-6 \theta_Z) ] \, , \\
a_3&=& 2 h^2 \mu^2 [1 + h^2 \mu^2 s_0 (5 s_0 - 2 \theta_Z)] \, ,\nonumber \\
a_4&=&h^4 \mu^4 (5 s_0 - \theta_Z) \, , \ \ a_5 = h^4 \mu^4 \, , \nonumber
\end{eqnarray}
where $\vartheta_{\pm}=1 \pm h^2 \mu^2 s_0^2$. Equation (21) is fifth
order and the corresponding solutions for $\Omega_Z^i(t+{\textstyle
\frac{h}{2}})$ are independent on parameter $s_0$, provided the unknown
$\delta$ is precisely determined. However, as is well known, only algebraic
equations of fourth or less orders allow to be solved in quadratures.

To overcome this difficulty, it is necessary to choose the parameter $s_0$
as a good prediction for $\Omega_Z^i(t+{\textstyle \frac{h}{2}})$ to
be entitled to ignore the highest-order terms in the left-hand side of
Eq.~(21). The simplest choice for this can be found assuming that the
nonlinear velocity term in the right-hand side of Eq.~(18) at $\alpha=Z$
is time-independent during the interval $[t-{\textstyle \frac{h}{2}},
t+{\textstyle \frac{h}{2}}]$, i.e., letting ${\Omega}_X^i(t+{\textstyle
\frac{h}{2}}) {\Omega}_Y^i(t+{\textstyle \frac{h}{2}})={\Omega}_X^i(t-
{\textstyle \frac{h}{2}}) {\Omega}_Y^i(t-{\textstyle \frac{h}{2}})+{\cal
O}(h)$. As a result, one obtains
\vspace{-2pt}
\begin{equation}
s_0=\theta_Z + h \rho_Z \Omega_X^i(t-{\textstyle \frac{h}{2}})
\Omega_Y^i(t-{\textstyle \frac{h}{2}})
\end{equation}

\vspace{-1pt}

\noindent
that represents the original values of $\Omega_Z^i(t+{\textstyle \frac{h}
{2}})$ with the second-order truncation error, so that $\delta={\cal O}
(h^2)$. It is easy to see that in this case the two last terms $a_4
\delta^4$ and $a_5 \delta^5$ in the left-hand side of Eq.~(21) behaves
as ${\cal O}(h^{12})$ and ${\cal O}(h^{14})$, respectively. Taking into
account the smallness of $h$, such terms can merely be omitted without
any loss of the precision, because they involve uncertainties of order
${\cal O}(h^{12})$ into the desired solution and appear to be too small
with respect to third-order truncation errors ${\cal O}(h^3)$ involved
initially in angular velocities by the algorithm.

Eq.~(21) is now transformed into the third-order algebraic equation
\begin{equation}
a_0+a_1 \delta+a_2 \delta^2+a_3 \delta^3={\cal O}(h^{12})
\end{equation}
which can easily be solved analytically,
\begin{eqnarray}
\delta_1&=&-{\textstyle \frac13} a_2/a_3+c-b/c +{\cal O}(h^{12}) \, , \\
\delta_{2,3}&=&-{\textstyle \frac13} a_2/a_3-{\textstyle \frac12}
[c-b/c \pm{\rm i} \sqrt{3} (c+b/c)]+{\cal O}(h^{12}) \, , \nonumber
\end{eqnarray}
where $b = {\textstyle \frac19}(3 a_1 a_3 - a_2^2)/a_3^2$, $c = (d +
\sqrt{b^3+d^2})^{1/3}$ and $d = {\textstyle \frac{1}{54}} (9 a_1 a_2 a_3
- 27 a_0 a_3^2 - 2 a_2^3)/a_3^3$. Among three solutions (25), only
the first one $\delta_1$ is real and satisfies the physical boundary
condition $\delta_1 \to h^2$ when $h$ goes to zero (the other two solutions
$\delta_{2,3}$ are purely imaginary at $h \to 0$ and they tend to infinity
as $\sim \pm {\rm i}/h$). Thus, the desired $Z$-th component of the angular
velocity is
\begin{equation}
\Omega_Z^i(t+{\textstyle \frac{h}{2}})=s_0+\delta_1 \, .
\end{equation}
The rest two components $\Omega_X^i(t+{\textstyle \frac{h}{2}})$ and
$\Omega_Y^i(t+{\textstyle \frac{h}{2}})$ are reproduced on the basis
of equalities (20).

\subsection{Symplectic properties}

It can be checked readily that past and future values of all the integrated
quantities enter symmetrically into Eqs. (5)--(7), (10), (14)--(17) and
(19). Therefore, the algorithm derived is time reversible with respect
to translational and rotational motions and within both microcanonical
and canonical ensembles. In order to investigate symplectic properties,
it is necessary to choose arbitrary canonically conjugated coordinates,
express them in terms of ${\bf r}_i$, ${\bf v}_i$, ${\bf A}_i$ and ${\bf
\Omega}_i$, and, then look whether the corresponding volume in phase space
is conserved. We accept the positions ${\bf r}_i^a={\bf r}_i+{\bf A}_i^+
{\bf \Delta}_a$ and momenta ${\bf p}_i^a = m_a({\bf v}_i + {\bf A}_i^+ \,
[{\bf \Omega}_i {\mbox{\boldmath $\times$}} {\mbox{\boldmath $\Delta$}}^a])$
of atomic sites to be such canonical coordinates.

Since the translational and rotational variables are not coupled explicitly
during our integration (in the microcanonical ensemble), the transitions
from old to new values of the canonical coordinates, caused by varying
these variables, can be considered separately at each time step. As is well
known, the translational leapfrog algorithm (5) is symplectic, so that the
corresponding transition will be performed with the unit Jacobian. In its
turn, the time evolution of orientational variables can be split into two
consequent transformations. During the first one, the principal angular
velocities ${\bf \Omega}_i$ are changed provided the orientational positions
${\bf A}_i$ remain constant, whereas the second transformation will
correspond to the change of ${\bf A}_i$ at fixed values of ${\bf \Omega}_i$
(a similar splitting is often used [46] to prove the symplecticity of usual
schemes). As far as the orientational matrices ${\bf A}_i$ appear to be
always orthonormal (${\rm det} {\bf A}_i=1$), the effect of their changes
in time is reduced simply to a rotation of vectors ${\bf r}_i^a-{\bf r}_i$
and ${\bf p}_i^a-m_a{\bf v}_i$ in three-dimensional space. Thus, the second
transformation is evidently volume preserving. Finally, since angular
velocities ${\bf \Omega}_i$ enter linearly into momenta ${\bf p}_i^a$, it
can be shown that the first transformation will conserve the volume too
provided the Jacobian ${\cal J}_{{\bf \Omega}_i}={\rm det} {\bf \Theta}_i$
is equal to unity, where ${\bf \Theta}_i=\partial \{ \Omega_X^i({t+
{\textstyle \frac{h}{2}}}), \Omega_Y^i({t+{\textstyle \frac{h}{2}}}),
\Omega_Z^i({t+{\textstyle \frac{h}{2}}})\}/\partial \{ \Omega_X^i({t-
{\textstyle \frac{h}{2}}}), \Omega_Y^i({t-{\textstyle \frac{h}{2}}}),
\Omega_Z^i({t-{\textstyle \frac{h}{2}}})\}$.

Partially differentiating each equation ($\alpha=X,Y,Z$) of system (6)
consequently over $\Omega_X^i({t-{\textstyle \frac{h}{2}}})$, $\Omega_Y^i
({t-{\textstyle \frac{h}{2}}})$ and $\Omega_Z^i({t-{\textstyle \frac{h}
{2}}})$, and solving the obtained three systems of linear equations with
respect to nine elements of ${\bf \Theta}_i$, one finds
\begin{equation}
{\cal J}_{{\bf \Omega}_i}=\frac{[1-h^2\{\sigma_Y \sigma_Z {\Omega_X^i}^2+
\sigma_X \sigma_Z {\Omega_Y^i}^2+\sigma_X \sigma_Y {\Omega_Z^i}^2\} + 2 h^3
\sigma_X \sigma_Y \sigma_Z \Omega_X^i \Omega_Y^i \Omega_Z^i]_{t-h/2}}
{[1-h^2\{\sigma_Y \sigma_Z {\Omega_X^i}^2+\sigma_X \sigma_Z {\Omega_Y^i}^2+
\sigma_X \sigma_Y {\Omega_Z^i}^2\} - 2 h^3 \sigma_X \sigma_Y \sigma_Z
\Omega_X^i \Omega_Y^i \Omega_Z^i]_{t+h/2}} \, .
\end{equation}
Therefore, unless $J_X=J_Y=J_Z$, the single discrete step within our
algorithm is not volume conserving, i.e., ${\cal J}_{{\bf \Omega}_i}(t)=
1+{\cal O}(h^3)$. However, since principal angular velocities are Gaussian
distributed in equilibrium with $\langle \Omega_X^i \Omega_Y^i \Omega_Z^i
\rangle=\langle \Omega_X^i \rangle \langle \Omega_Y^i \rangle \langle
\Omega_Z^i \rangle = 0$, the overall Jacobian ${\cal J}_{\cal N} =
\prod_n^{\cal N} {\cal J}_{{\bf \Omega}_i}(t+n h)$ will fluctuate around
unity. Moreover, the probability of deviations of ${\cal J}_{\cal N}$ from
unity on some value $u$ will decrease rapidly (as $1/\exp(\sim u/h^3)$)
with increasing $u$ and will not depend on the total time step ${\cal N}$
performed. In other words, such deviations are not accumulated during the
integration.

From the afore said, it becomes clear that to reduce the fluctuations
of ${\cal J}_{{\bf \Omega}_i}(t)$ to zero, it is necessary to take
into account the contributions $(J_\beta-J_\gamma) {\Omega}_\beta^i
{{\Omega}_\gamma^i} \equiv [{\bf W}({\bf \Omega}_i) {\bf J} {\bf
\Omega}_i]_\alpha$ of free-motion torques into the angular-velocity
dynamics more precisely (here ${\bf J}$ denotes the diagonal matrix of
principal moments of inertia). To do this, let us write the solutions
${\bf \Omega}_i({t+{\textstyle \frac{h}{2}}})$ to Eq.~(6) in the
following formal form ${\hat {\cal S}}(h) {\bf \Omega}_i({t-{\textstyle
\frac{h}{2}}})$, where ${\hat {\cal S}}(h)$ is a velocity-displacement
operator. It is obvious by construction that $\lim_{s \to \infty} [{\hat
{\cal S}}(h/s)]^s = {\rm e}^{L h}$, where $L={\bf J}^{-1} [{\bf K}_i +
{\bf W}({\bf \Omega}_i) {\bf J} {\bf \Omega}_i] \partial/\partial {\bf
\Omega}_i$ is a part of the Liouville operator of the system, so that
${\rm d} {\bf \Omega}_i/{\rm d} t=L {\bf \Omega}_i$. Of course, the
evolution operator ${\rm e}^{L h}$ does not lead, in general, to exact
solutions, since the torque caused by potential forces is assumed to
be constant, ${\bf K}_i \equiv {\bf K}_i(t)$, over the time interval
$[t-{\textstyle \frac{h}{2}},t+{\textstyle \frac{h}{2}}]$. Thus like
${\hat {\cal S}}(h)$, the propagator ${\rm e}^{L h}$ generates new
angular velocities with the same ${\cal O}(h^3)$ order local errors.
However, the operator $L$ allows to be decomposed into a sum $L_1+L_2$,
and then the (Trotter) formula ${\rm e}^{L h}={\rm e}^{L_2 {\textstyle
\frac{h}{2}}} {\rm e}^{L_1 h} {\rm e}^{L_2 {\textstyle \frac{h}{2}}} +
{\cal O}(h^3)$ can be used to approximate the full propagator. Although
different decompositions are possible, it is essential for our purposes,
to decompose $L$ in such a way to solve the problem analytically.

The last requirement leads to the following decomposition
\begin{equation}
L_1={\bf J}^{-1}
{\bf K}_i \frac{\partial}{\partial {\bf \Omega}_i} \, , \ \ \ \ \ \ \
L_2={\bf J}^{-1} {\bf W}({\bf \Omega}_i) {\bf J} {\bf \Omega}_i
\frac{\partial}{\partial {\bf \Omega}_i} \, .
\end{equation}
Then ${\rm e}^{L_1 h} {\bf \Omega}_i={\bf \Omega}_i + h {\bf J}^{-1} {\bf
K}_i$, whereas the propagator ${\rm e}^{L_2 h/2}$ corresponds to a free
rotational dynamics. As is well known, Euler equations (4) are integrated
analytically in this case. Let $\frac12 \sum_{\alpha}^{X,Y,Z} J_\alpha
{\Omega_\alpha}^2={\cal H}$ and $\frac12 \sum_{\alpha}^{X,Y,Z} {J_\alpha}^2
{\Omega_\alpha}^2={\cal M}^2$ be the kinetic energy and square angular
momenta of a body, associated with angular velocity ${\bf \Omega}$,
$j_\alpha={\textstyle \frac12}(-\sigma_\beta \sigma_\gamma)^{-1/2}$,
$\varepsilon=[(J_Z-J_Y)({\cal M}^2 -2 {\cal H} J_X)/J_X J_Y J_Z]^{1/2}$
and $\kappa=[(J_Y-J_X)(2 {\cal H} J_Z-{\cal M}^2)/((J_Z-J_Y)({\cal M}^2
-2{\cal H} J_X))]^{1/2}$. Then, the result for transformed angular
velocities ${\bf \Omega'}={\rm e}^{L_2 h/2} {\bf \Omega}$ can be
presented in the form [47]:
\vspace{-2pt}
\begin{eqnarray}
\Omega_X'&=&\kappa j_X \varepsilon {\rm cn}
(\varepsilon {\textstyle \frac{h}{2}}+\upsilon) \, , \nonumber \\
\Omega_Y'&=&{\rm sign}(\Omega_Z) \kappa j_Y \varepsilon {\rm sn}
(\varepsilon {\textstyle \frac{h}{2}}+\upsilon) \, , \\
\Omega_Z'&=&{\rm sign}(\Omega_Z) j_Z \varepsilon {\rm dn}
(\varepsilon {\textstyle \frac{h}{2}}+\upsilon) \, , \nonumber
\end{eqnarray}

\vspace{-2pt}

\noindent
where ${\rm sign}(x)$ denotes sign of $x$, ${\rm dn}(x)=[1-\kappa^2
\sin^2({\rm am}(x))]^{1/2}$, ${\rm sn}(x)=\sin({\rm am}(x))$ and
${\rm cn}(x)=\cos({\rm am}(x))$ are Jacobian elliptic functions [48],
so that $\Xi({\rm am}(x),\kappa)=\int_0^{{\rm am}(x)} {\rm d} \psi/
[1-\kappa^2 \sin^2\psi]^{1/2}=x$ with $\Xi$ being a Legendre elliptic
integral of the first kind (the program code for these functions can be
found in [49]). The constant $\upsilon$ is determined from the initial
boundary condition $\lim_{h \to 0} {\bf \Omega'}={\bf \Omega}$, which
leads to $\upsilon= \overline{\upsilon} \, {\rm sign}(\Omega_Y \Omega_Z
{\rm sn}(\overline{\upsilon}))$, where $\overline{\upsilon}=\Xi(\arccos
(\Omega_X/(j_X \kappa \varepsilon)),\kappa)$. Note that analytical
solutions (29) are applicable when ${\cal M}^2 \ge 2 {\cal H} J_Y$.
Otherwise, the indexes $X$ and $Z$ should be replaced between themselves
in Eq.~(29) and in expressions for $\varepsilon$, $\kappa$ and $\upsilon$.

It can be checked directly that the Jacobian of transformation ${\bf
\Omega'}={\rm e}^{L_2 h/2} {\bf \Omega}$ is equal to unity. This valid
also for the operator ${\rm e}^{L_1 h}$, because a simple shift does not
change the volume. Therefore, if the angular velocities are integrated,
instead of (6), as
\begin{equation}
{\bf \Omega}_i({t+{\textstyle \frac{h}{2}}})=
{\rm e}^{L_2 {\textstyle \frac{h}{2}}} {\rm e}^{L_1 h}
{\rm e}^{L_2 {\textstyle \frac{h}{2}}}
{\bf \Omega}_i({t-{\textstyle \frac{h}{2}}}) \, ,
\end{equation}
our algorithm will be symplectic in rigorous meaning.

\subsection{Integration in specific cases}

Although the algorithm described can be applied to arbitrary rigid
polyatomics, some simplifications are possible using special properties
of the molecule. The simplest case is molecules with a spherical distribution
of mass, when $J_X=J_Y=J_Z \equiv J$. Then, no free-motion torques appear,
and it is more convenient to work within the body-vector representation and
to rewrite equations (1) and (4) in terms of angular velocities ${\mbox
{\boldmath $\omega$}}_i={\bf A}_i^+{\bf \Omega}_i$ in the laboratory frame,
i.e., ${\rm d} {\bf A}_i/{\rm d} t={\bf A}_i {\bf W}({\mbox{\boldmath
$\omega$}}_i)$ and $J {\rm d} {\mbox{\boldmath $\omega$}}_i/{\rm d} t={\bf
k}_i -\lambda J {\mbox{\boldmath $\omega$}}_i$. The leapfrog trajectories
for these equations are obvious: ${\mbox{\boldmath $\omega$}}_i(t+\frac{h}
{2})=[\nu_-(t) {\mbox{\boldmath $\omega$}}_i(t-\frac{h}{2})+\frac{h}{J}\,
{\bf k}_i(t)]/\nu_+(t)$ and ${\bf A}_i(t+h)={\bf A}_i(t) {\rm \bf exp}
[\varphi_i {\bf W}({\mbox{\boldmath $\omega$}}_i)/\omega_i]_{t+\frac{h}
{2}}$, where $\varphi_i=\arcsin [h \omega_i/(1+{\textstyle \frac{h^2}{4}}
\omega_i^2)]_{t+\frac{h}{2}}$.

For some particular models, the orientational part of intermolecular
potentials can be expressed using only unit vectors ${\mbox{\boldmath
$\rho$}}_i$ passing through the centers of mass of molecules. The examples
are point dipole interactions, when ${\mbox{\boldmath $\rho$}}_i \equiv
{\mbox{\boldmath $\nu$}}_i/\nu_i$ with ${\mbox{\boldmath $\nu$}}_i$ being
the dipole moment, or when all force sites of the molecule are aligned
along ${\mbox{\boldmath $\rho$}}_i$. If then additionally the condition
$J_{X,Y,Z}=J$ is satisfied (for the last example this can be possible when
forceless mass sites are placed in such a way to ensure this condition), it
is no longer necessary to deal with orientational matrices or quaternions.
In this case the equation for ${\mbox{\boldmath $\rho$}}_i$ looks as ${\rm
d} {\mbox{\boldmath $\rho$}}_i/{\rm d} t={\bf W}^+({\mbox{\boldmath
$\omega$}}_i) {\mbox{\boldmath $\rho$}}_i$ with the solution ${\mbox
{\boldmath $\rho$}}_i(t+h)={\rm \bf exp}[-\varphi_i {\bf W}({\mbox
{\boldmath $\omega$}}_i)/\omega_i]_{t+\frac{h}{2}} {\mbox{\boldmath
$\rho$}}_i(t)$.

For molecules with the cylindric symmetry of mass distribution, the
numerical trajectory can also be determined in a simpler manner. Let us
assume for definiteness that $J_X=J_Y \ne J_Z \ne 0$. Then arbitrary two
perpendicular between themselves axes, lying in the plane perpendicular
to $Z$-th principal axis, can be chosen initially as $X$- and $Y$-th
principal orths. The corresponding solution to Eq.~(17) at $\alpha=Z$
is found now exactly, namely, $\Omega_Z^i(t+{\textstyle \frac{h}{2}})=
[\nu_-(t) \Omega_Z^i(t-{\textstyle \frac{h}{2}})+\frac{h}{J_Z} K_Z^i(t)]/
\nu_+(t)$ (the $X$- and $Y$-th components are obtained automatically in
view of Eq.~(20)), whereas the orientational matrices or quaternions
are computed via Eq.~(11).

A special attention should be paid on linear molecules when $J_X=J_Y=J
\ne J_Z=0$. Each such molecule has two orientational degrees of freedom
and to reproduce a correct dynamics by Euler equations it is necessary to
putt formally $\Omega_Z^i \equiv 0$ to exclude nonexisting torques caused
by irrelevant rotations of the molecule around $Z$-axis. Then, it follows
from Eq.~(20) that $\Omega_{X,Y}^i(t+{\textstyle \frac{h}{2}})=[\nu_-(t)
\Omega_{X,Y}^i(t-{\textstyle \frac{h}{2}})+\frac{h}{J} K_{X,Y}^i(t)]/
\nu_+(t)$. Planar molecules do not present a specific case within our
approach and they are handled as tree-dimensional bodies.

\section{MD tests. Comparison with previous methods}

The system chosen was the TIP4P model of water [50] at a density of $m N/V$=
1 g/cm$^3$ and temperature of $T$=298 K. Such a system should provide a very
severe test for rotational algorithms because of the low moments of inertia
of the molecule and the large torques due to the site-site interactions. To
reduce cut-off effects we used a cubic sample of $N=256$ molecules and the
reaction field geometry [51]. Our MD programs were implemented using Fortran
language and double precision throughout. They were executed on a Pentium-S
120 MHz personal computer at around 0.8 s per time step. All runs were
started from an identical well equilibrated configuration.

We have made comparative tests on the basis of our advanced leapfrog
algorithm, the implicit leapfrog algorithm of Fincham [26], pseudo-site
formalism [23], angular-velocity Verlet method within matrix- and
quaternion-constraint schemes [27, 28], and the atomic-constraint
technique [17, 18]. The results obtained for the total energy fluctuations
${\cal E}=[\langle (E-\langle E \rangle)^2 \rangle ]^{1/2}/|\langle E
\rangle|$ as functions of the length (${\cal N}=t/h$) of the microcanonical
(NVE) simulations are shown in Fig.~1 at four fixed step sizes, $h=$ 1, 2,
3 and 4 fs (a step size of 2 fs is normally used [52, 53] to simulate water
within the atomic-constraint approach). As can be seen, the Fincham's
leapfrog (marked simply as ''leapfrog`` in Fig.~1) and pseudo-site schemes
appear to be unstable already for the least time step considered and they
lead to the worst energy conservation. Much more stable trajectories
are produced by the velocity-Verlet integrator within quaternion- and
matrix-constraint schemes which exhibit similar equivalence in the energy
conservation. But the results are rather poor at moderate and great time
steps ($h \ge$ 3 fs). Only the atomic-constraint and our advanced leapfrog
algorithms can be related to long-term stable schemes.

We mention that the computation of total energy $E={\mit \Gamma}+U$
(consisting of kinetic ${\mit \Gamma}$ and potential $U$ parts) at time
$t$ within the leapfrog framework requires the knowledge of on-step
velocities. These velocities can be calculated using usual estimators
(14) and (16). The corresponding curves of dependencies ${\cal E}({\cal
N})$ are labeled by ''1`` in Fig.~1. They are almost identical to those
obtained within the atomic-constraint technique. Note that ${\cal O}(h^2)$
uncertainties, arising in evaluations (14) and (16), are in the
self-consistency with second-order global errors appearing during the
leapfrog integration. Despite this, an additional portion is involved
into the main ${\cal O}(h^2)$ term of accumulated errors for ${\cal E}$,
increasing the total energy fluctuations with no relation to the real
accuracy of the computed trajectory [54]. Although various more accurate
estimators are available [55], we have established that the following
four-point symmetric scheme
\begin{equation}
{\bf V}(t)=\frac{1}{16} \Big[
-{\bf V}(t-{\textstyle \frac{3h}{2}})
+9{\bf V}(t-{\textstyle \frac{h}{2}})
+9{\bf V}(t+{\textstyle \frac{h}{2}})
- {\bf V}(t+{\textstyle \frac{3h}{2}}) \Big] + {\cal O}(h^4) \, ,
\end{equation}
where ${\bf V} \equiv \{ {\bf v}_i, {\bf \Omega}_i \}$, leads to the best
energy conservation within our leapfrog algorithm. It is understood, of
course, that mid-step velocities entering into the right-hand site of
Eq.~(31) are already defined quantities. Thus, the computation of the
total energy $E$ at time $t$ becomes possible, when the velocity step
$t+{\textstyle \frac{3h}{2}}$ has been passed and the potential energy
$U(t)$ (known at this stage already at $t+h$) has been taken from memory.
The corresponding dependencies of ${\cal E}$ on ${\cal N}$ are plotted in
Fig.~1 by the boldest (lowest lying) curves marked as ''2``. In this case,
the total-energy fluctuations decrease about in 1.5 times with respect to
the usual two-point scheme.

The quaternion and principal-axis representations of the advanced leapfrog
algorithm conserved the energy approximately with the same accuracy. For
this reason, the result concerning principal-axis variables is not plotted
in Fig.~1 to simplify the graph. The leapfrog trajectories were generated
applying quasianalytical solutions (Eqs.~(20) and (26)) for angular
velocities. The solutions obtained by means of iterations of Eq.~(6) were
calculated also for comparison. No deviations between the both results have
been identified up to $h=6$ fs. They differed on each step by uncertainties
of order round-off errors only, so that the free-of-iteration scheme appears
to be in an excellent accord. No shift of the total energy and temperature
was observed during the advanced leapfrog trajectories at $h \le 5$ fs over
a length of 10 000 time steps. The deviations from unity of the overall
Jacobian ${\cal J}_{\cal N}$ (see Eq.~(27)) never exceed about 5\% (at $h
\le 4$ fs).

Finally, the rigorously symplectic version (30) has also been examined
(the longest-dashed curves in Fig.~1) within the four-point scheme (31).
As we can see, this version does not lead to improvements in energy
conservation, despite the fact that the free-motion propagator ${\rm
e}^{L_2 h/2}$ is evaluated exactly. This is so because for water at the
given thermodynamic point, the free-motion torques are much smaller in
amplitude with respect to the torques caused by interactions. An increased
conservation of energy can be expected for systems at high temperatures,
where the free-motion contributions into the torques become dominate.
Otherwise, the analytical version (29) is not generally recommended because
it requires the calculation of somewhat time-consuming elliptic functions
(although we did not observe any considerable decreasing time efficient,
given that near 95\% of the total computer time were spent to evaluate
pair interactions). In the limiting case when the potential-force torques
are absent at all, symplectic solutions (29) will lead to exact results
with automatic preservation of energy and angular momenta. The symplectic
version should also be used in situations where a precise conservation of
the volume in phase space at each time step is very important. It is worth
remarking that the variant $L_1 \leftrightarrow L_2$ of decomposition (28)
in transformation (30) is also acceptable in view of the symplecticity. We
have established, however, that such a decomposition leads to worse energy
conservation in our simulations (probably because of $\langle {{\bf K}_i}^2
\rangle \gg \langle [{\bf W}({\bf \Omega}_i) {\bf J} {\bf \Omega}_i]^2
\rangle$).

To properly reproduce features of NVE ensembles, it is necessary for the
ratio $\Upsilon$ of ${\cal E}$ to the fluctuations ${\cal U}$ of potential
energy to be no more than a few per cent. We have obtained the following
levels of ${\cal E}$ (within the two-point scheme) at the end of these
trajectories: 0.0016, 0.0065, 0.015, 0.029, 0.049 and 0.10 \%, corresponding
to $\Upsilon \approx$ 0.29, 1.2, 2.7, 5.2, 8.7 and 18 \% at $h=$ 1, 2, 3, 4,
5 and 6 fs, respectively (${\cal U} \approx 0.56 \%$ for the system under
consideration). Therefore, a step size of 4 fs is still suitable for
precise calculations. The greatest time steps 5 fs and 6 fs can sometimes
be acceptable when the precision is not so important, for example, for the
equilibration of configurations. The ratio $\Upsilon$ in the interval $h
\le 5$ fs can be fitted with a great accuracy to the function $C h^2$ with
the coefficient $C \approx 0.29$ \% fs$^{-2}$. This is completely in line
with the square growth of global errors appearing during the integration
by Verlet-type integrators [54].

To verify the angular-velocity approach in more detail, in our NVE and
canonical (NVT) ensemble simulations we measured besides the total energy
and temperature some other relevant functions of the system, namely,
specific heat at constant volume, mean-square forces and torques,
oxygen-oxygen and hydrogen-hydrogen radial distribution functions (RDFs).
Center-of-mass (CM) and angular-velocity (AV) time autocorrelation
functions (TAFs) were also found. Orientational relaxation was studied
by evaluating the molecular dipole-axis (DA) autocorrelations. The NVE
simulations, performed within the atomic-constraint technique at $h=$ 2
fs, was considered as a benchmark against which other algorithms and step
sizes are to be compared. First of all, to finish the discussion with the
NVE integrators, we report that deviations in all the measured functions
with respect to their benchmark values were in a complete agreement with
the corresponding relative fluctuations $\Upsilon$. For example, the
results obtained with the help of the advanced leapfrog algorithm at $h=$ 2
fs were indistinguishable from the benchmark ones. At the same time, they
differed as large as around 5\%, 10\% or even 20\% with increasing the
time step to 4 fs, 5 fs or 6 fs, respectively, however, these differences
were smaller than for other integrators. Therefore, there is a little point
in pursuing the energy-conserving algorithms to time steps larger than 4 fs
because the deviations become evident.

In the case of NVT simulations, the investigated quantities were less
sensitive to the step size increasing. These simulations have been performed
using the Fincham implicit algorithm [26] (within the Brown and Clarke
thermobath [37]) as well as the advanced leapfrog integrator within the
N\'ose-Hoover thermostat (Sec. II C) applying the free-of-iteration scheme
(Eqs.~(20) and (26)). The thermostat relaxation time $\tau$ was chosen to be
1 ps, (i.e., $h \ll \tau \ll {\cal N} h$) and the friction coefficient
$\lambda(t)$ was putted to be zero at the very beginning $(t=0)$ of the
simulations. The RDFs and CM, AV and DA TAFs calculated during the NVT
integrations at different step sizes ($h=$ 2--10 fs) are shown in Fig.~2
in comparison with the benchmark results. These functions at $h=4$ and 6
fs coincided completely with those corresponding to $h=2$ fs and they are
not shown in subsets (a)--(b) to simplify the presentation. As can be seen
from Fig.~2, the RDFs remain practically the same at $h \le 8$ fs in the
case of the advanced integrator. The deviations of all the correlation
functions from the benchmark within the usual implicit algorithm are
clearly larger. The orientational correlation function (see subset (d))
is to show a systematic discrepancy in this case. For instance, these
deviations at $h=$ 4 fs are as big as those obtained during the advanced
leapfrog integration at $h=$ 8 fs. Therefore, the last approach allows a
step size approximately twice larger than the usual implicit integrator.
We conclude, therefore, that the thermostatted advanced leapfrog integrator
allows to be used up to a time step of 6 fs, given that then there is no
difference in RDFs, while CM, AV and DA TAFs are also close.

\section{Concluding remarks}

We have formulated a new approach for numerical integration of the
equations of motion for systems with interacting rigid bodies. Unlike
other standard methods, the principal angular velocities are involved
directly into the integration within this approach. The algorithm derived
is categorized as a rotational leapfrog, since the variables saved are
mid-step angular velocities and on-step orientational positions. The
orientations can be expressed in terms of either quaternions or entire
rotational matrices. The interpolation of velocity- and orientation-dependent
quantities to the corresponding middle time points was carried out using a
simple averaging over the two nearest neighboring values that is in the
spirit of the leapfrog idea as well. As a result, the following significant
benefits have been achieved: (i) the exact conservation of rigid structures
appears to be an intrinsic feature of the algorithm, and (ii) all the
evaluations are performed analytically in both NVE and NVT ensembles
without involving any iterative procedures.

It has been shown on the basis of an actual computer experiment on water
that the algorithm presented exhibits better energy conserving properties
than those observed in all other rigid-motion integrators known. The
algorithm can easily be implemented for arbitrary rigid bodies and
substituted into existing program codes. It seems to be a good alternative
to the atomic-constraint method in the case of long term MD simulations of
systems with rigid polyatomic molecules.

The approach introduced can be developed to perform a MTS integration
within the RESP technique. This question will be considered in our further
studying. Some ideas of the MTS integration have already be used in this
paper to derive symplectic versions of the advanced leapfrog algorithm. The
reformulation of the RESP approach in generalized coordinates containing
translational and orientational variables explicitly may lead to significant
simplifications when selecting efficient reference system propagators. For
example, translational motions (which are much slower than rotational ones
for the most of liquids) can be decomposed directly within such coordinates.
The most notorious demonstration is very fast rotations of rigid bodies
in the inertial-motion regime. Then for molecules with a spherically
symmetric distribution of mass we merely obtain ${\bf v}_i=$ const and
${\bf \Omega}_i=$ const within the molecular approach. To reproduce such
a behavior within the atomic technique, it is necessary to integrate (with
a very small step size) the equations of motion in the presence of rapidly
varying strong constraint forces. The decomposition of rotational
propagators containing interactions into slow and fast parts can also
be done easily splitting in an appropriate way the free-motion and
potential-force torques.

\vspace{12pt}

{\bf Acknowledgements.} This work was financially supported in part by a
grant of the President of Ukraine. I wish to thank Dr. A. Duviryak for
helpful discussions. I would like also to thank Professor B.J. Berne and
Professor G.J. Martyna for sending me reprints of some articles.

\newpage

\small
\addtolength{\baselineskip}{-5.75pt}

\addtolength{\baselineskip}{5.75pt}
\normalsize

\newpage

\vspace*{1cm}

\begin{center}
{\large Figure captions}
\end{center}

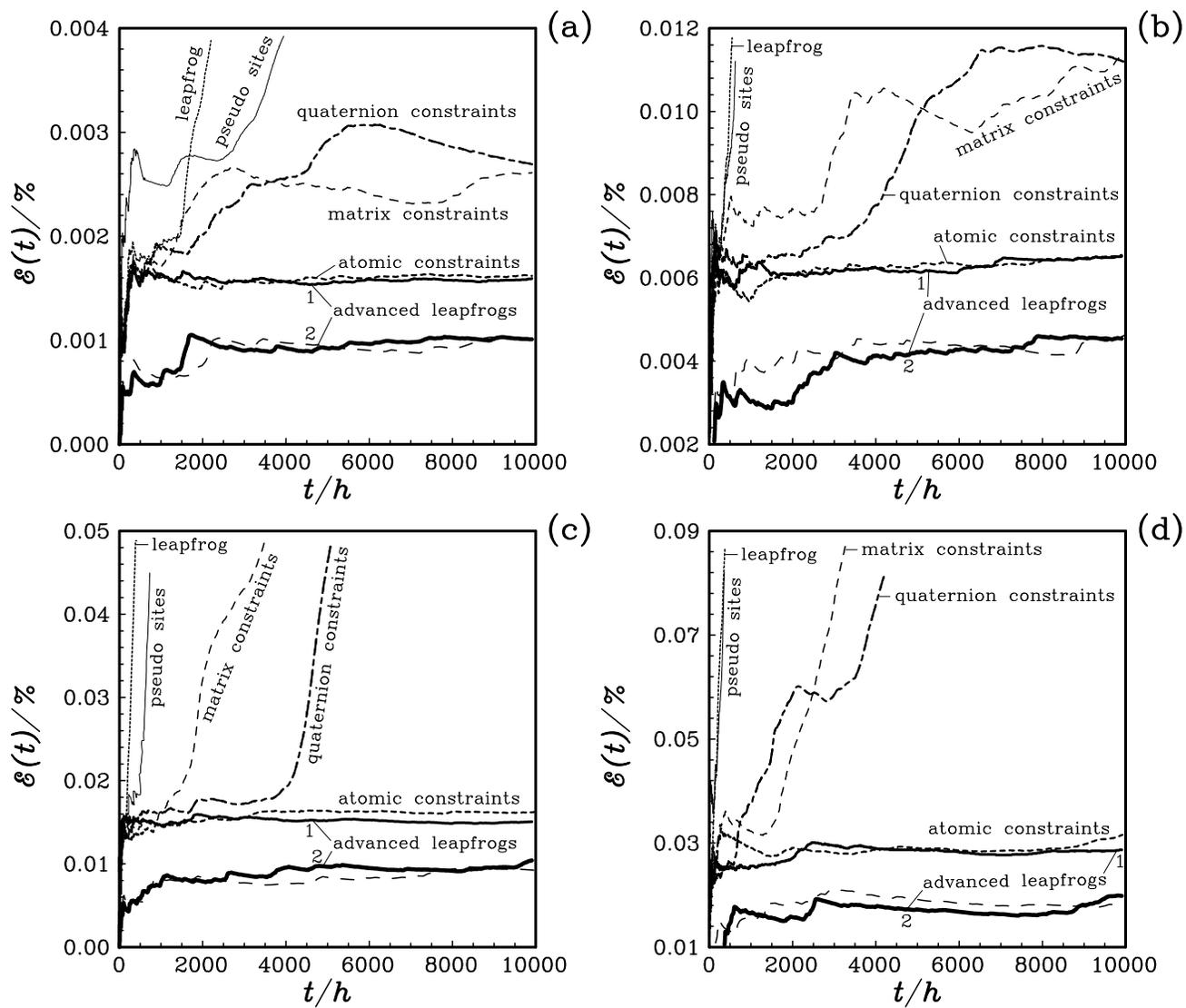
Fig. 1. The total energy fluctuations as functions of the length of the
NVE simulations on the TIP4P water, performed in various techniques at
four fixed time steps: {\bf (a)} 1 fs, {\bf (b)} 2 fs, {\bf (c)} 3 fs
and {\bf (d)} 4 fs.

\vspace{12pt}

Fig. 2. Oxygen-oxygen (O-O) and hydrogen-hydrogen (H-H) radial distribution
functions {\bf (a)}, center-of-mass {\bf (b)} and angular-velocity {\bf (c)}
time autocorrelation functions, and orientational relaxation {\bf (d)},
obtained in the NVT simulations of the TIP4P water. The results corresponding
to the step sizes $h=$ 2, 8 and 10 fs are plotted by bold solid, short-dashed
and thin solid curves, respectively. Additional long-short dashed and dashed
curves in {\bf (c)}--{\bf (d)} correspond to the cases of $h=$ 4 and 6 fs.
The sets of curves related to the usual implicit and advanced leapfrog
algorithms are labeled by "U" and "A", respectively (the result of the
usual implicit algorithm is not included in {\bf (c)} to simplify the
graph). The benchmark data are shown as open circles. Note that the
advanced-algorithm curves are indistinguishable in {\bf (d)} at
$h=$ 2, 4 and 6 fs.

\end{document}